# Overview of EIREX 2012: Social Media

Julián Urbano, Mónica Marrero, Diego Martín and Jorge Morato

University Carlos III of Madrid
Department of Computer Science
Avda. Universidad, 30
Leganés, Madrid, Spain

## 1. Introduction

The third Information Retrieval Education through EXperimentation track (EIREX 2012) was run at the University Carlos III of Madrid, during the 2012 spring semester.

EIREX 2012 is the third in a series of experiments designed to foster new Information Retrieval (IR) education methodologies and resources, with the specific goal of teaching undergraduate IR courses from an experimental perspective. For an introduction to the motivation behind the EIREX experiments, see the first sections of [Urbano et al., 2011a]. For information on other editions of EIREX and related data, see the website at http://ir.kr.inf.uc3m.es/eirex/.

The EIREX series have the following goals:

- To help students get a view of the Information Retrieval process as they would find it in a real-world scenario, either industrial or academic.
- To make students realize the importance of laboratory experiments in Computer Science and have them initiated in their execution and analysis.
- To create a public repository of resources to teach Information Retrieval courses.
- To seek the collaboration and active participation of other Universities in this endeavor.

This overview paper summarizes the results of the EIREX 2012 track, focusing on the creation of the test collection and the analysis to assess its reliability. Next section provides a brief overview of our course and the student systems. Section 3 describes the process we followed to create the EIREX 2012 test collection, and Section 4 presents the evaluation results. Section 5 analyzes the reliability of our approach by studying the effect of the incompleteness of judgments. Section 6 wraps up with the conclusions.

## 2. Teaching Methodology

EIREX 2012 took place during the 2012 spring semester, in the context of the Information Retrieval and Access course [Urbano et al., 2010b], which is an elective course taken by senior Computer Science undergraduates. In this course we teach traditional IR techniques, and the main lab assignment consists in the development, from scratch, of a search engine for HTML documents. The development of this search engine is divided in three modules to hand in separately:

- Module 1 contains the implementation of an indexer for a collection of HTML documents and a simple retrieval model for automatic ad hoc queries.
- Module 2 incorporates query expansion in the retrieval process.
- Module 3 adds simple Named Entity Recognition (NER) capabilities to aid in the "who" questions.

In this edition we had 70 students, who created a total of 22 systems in groups of 4 or 3 students each. Thus, we had a total of 66 systems: 22 with basic retrieval, 22 with query expansion and 22 with NER. However, due to several reasons some students dropped from the course; in the end we had 19 groups submitting a total of 56 systems. We try to encourage students by giving one extra point to the group who developed the most effective search engine (see Section 4).

All these systems are evaluated with an IR test collection built with the students also from scratch (see Section 3). A test collection for Information Retrieval evaluation contains three major components: a document collection, a set of information needs (usually called topics), and the relevance judgments or ground truth (usually assessed by humans) telling what documents are relevant to the topics [Voorhees, 2002]. The students run their systems for each topic, returning the list of documents in the collection deemed relevant to it. Then, we use some effectiveness measures to assess, according to the relevance judgments, how well the systems actually answered the information needs. This provides us with a ranking of the student systems in terms of effectiveness.

This year, we provided students with the full EIREX 2010 [Urbano et al., 2011b] and EIREX 2011 [Urbano et al., 2012] test collections to develop and train their systems. We then evaluated them with the EIREX 2012 test collection and published the results. The same process was repeated for the three modules of each student system.

## 3. Test Collection

The process we employ to create the EIREX test collections is different from those usually followed in other IR evaluation workshops such as the early ad hoc tracks of the Text REtrieval Conference (TREC) ran by NIST [Voorhees et al., 2005]. Although we follow very similar principles, working with students bears some limitations, mainly in terms of effort and reliability. The document collection cannot be as large as those usually employed in TREC, because undergraduate students do not have the adequate expertise to handle that much information and they would probably dedicate too much time to efficiency issues rather than effectiveness and the implementation and understanding of the IR techniques we explain. This limitation restricts the topics to use: if they had nothing in common, we would probably need too many documents to have sufficient diversity to include relevant material for each topic; but if they were somehow similar, probably fewer documents would be needed. Therefore, we decided that all topics should have a common theme, which in addition reflects more closely a real setting where students have to deploy an IR system for a company in a particular domain. Thus, the document collection depends on the topics and not the other way around as usual. For this third EIREX edition we chose the theme to be *Social Media*, as it is a topic of recent interest in Computer Science research.

The problem at this point is how to build a document collection making sure that some relevant material is included for every topic. In EIREX 2010 the topics were chosen by the course instructors alone, but in EIREX 2011 we involved students in the topic creation too; we did this also this year. Each student had to come up with three candidate topics about social media. For each topic, they had to issue queries to Google Web Search just as if they were trying to satisfy the information needs themselves, manually using term proximity operators, query expansion, etc.; and report an estimation of the amount of relevant material found in the first page of results. Based on this information we discarded topics apparently too difficult, with very few documents, or for which there did not seem to be clearly relevant documents. Once the final topic set was established, we used a focused web crawler to download all web pages returned by Google for each topic [Urbano et al., 2010a]. The union of all these web pages conform the *complete* document collection.

At this point we have a document collection and a set of topics, so next we need relevance judgments. Another difference here is that students have to make all relevance judgments *before* they start developing, as otherwise some might try cheating and judge all documents retrieved by their system as relevant. In addition, having them inspect the documents to assess their relevance helps later on during development because they know what kind of documents their systems will have to handle. Judging every document for every topic is completely impractical because it requires too much effort, so instead a sample of documents is judged for each topic (i.e. the topics' pools). To come up with a reliable pool of documents despite student systems do not directly contribute, we use well-known and freely available IR tools instead: Lemur[1] and Lucene[2] (call these the *pooling* systems). We thus proceed to index the complete document collection and obtain the results provided by different configurations of the pooling systems for each of the topics, trying to exploit as much as possible our previous knowledge about the topic and the information documents must contain to be considered relevant. For instance, if the topic asked for information about the CEO of a company, we would include the name of the person in the query. Unlike in 2010 and 2011, this year we also included Google as a pooling system, so that the top documents that Google returned for each topic would also be included.

---

[1] http://www.lemurproject.org
[2] http://lucene.apache.org

| Topic | Downloaded | Complete → Pooled | | Pooled → Biased | |
|---|---|---|---|---|---|
| | | Pool size | Pool depth | Pool size | Pool depth |
| 2012-001 | 638 | 102 | 28 | 162 | 113 |
| 2012-002 | 699 | 100 | 30 | 163 | 79 |
| 2012-003 | 869 | 105 | 20 | 161 | 70 |
| 2012-004 | 854 | 101 | 22 | 160 | 87 |
| 2012-005 | 816 | 101 | 29 | 161 | 110 |
| 2012-006 | 504 | 101 | 16 | 161 | 87 |
| 2012-007 | 667 | 101 | 33 | 160 | 111 |
| 2012-008 | 602 | 104 | 19 | 162 | 55 |
| 2012-009 | 643 | 100 | 29 | 160 | 119 |
| 2012-010 | 493 | 102 | 26 | 160 | 99 |
| 2012-011 | 827 | 103 | 27 | 160 | 96 |
| 2012-012 | 587 | 100 | 19 | 160 | 89 |
| 2012-013 | 339 | 100 | 28 | 160 | 97 |
| 2012-014 | 714 | 102 | 24 | 160 | 129 |
| 2012-015 | 557 | 103 | 26 | 164 | 118 |
| 2012-016 | 746 | 101 | 34 | 160 | 123 |
| 2012-017 | 824 | 104 | 22 | 163 | 92 |
| 2012-018 | 487 | 101 | 22 | 160 | 85 |
| 2012-019 | 441 | 101 | 31 | 161 | 95 |
| 2012-020 | 620 | 101 | 22 | 161 | 72 |
| 2012-021 | 644 | 101 | 21 | 160 | 105 |
| 2012-022 | 726 | 103 | 23 | 160 | 87 |
| 2012-023 | 507 | 104 | 22 | 161 | 102 |
| 2012-024 | 398 | 103 | 40 | 160 | 117 |
| 2012-025 | 558 | 105 | 21 | 160 | 118 |
| 2012-026 | 368 | 100 | 27 | 160 | 82 |
| 2012-027 | 756 | 104 | 22 | 160 | 66 |
| 2012-028 | 203 | 101 | 41 | 160 | 96 |
| 2012-029 | 669 | 101 | 22 | 161 | 54 |
| 2012-030 | 678 | 103 | 17 | 161 | 51 |
| 2012-031 | 711 | 102 | 24 | 160 | 98 |
| 2012-032 | 219 | 100 | 24 | 160 | 96 |
| 2012-033 | 654 | 100 | 36 | 160 | 122 |
| 2012-034 | 547 | 100 | 26 | 160 | 104 |
| 2012-035 | 668 | 103 | 22 | 161 | 50 |
| 2012-036[†] | 72 | - | - | - | - |
| 2012-037[†] | 317 | - | - | - | - |
| 2012-038[†] | 780 | - | - | - | - |
| Average | 590 | 102 | 26 | 161 | 94 |
| Total | 22,402 | 3,388 | - | 3,738 | - |

Table 1. Summary of the EIREX 2012 test collection. [†] for noise topics.

Pools are formed differently too: if we calculate *depth-k* pools (joining the top *k* results from the pooling systems), some topics might have considerably more documents to judge than others, as the final number depends on the overlap among the results. If some students were assigned a pool significantly larger than others, they could just judge carelessly once they think they have done enough work compared to their classmates. To prevent this situation we compute *size-k* pools instead: pools with the minimum depth such that the total size is at least *k* documents. Thus, each topic has a pool of documents with different depth, although all pools have very similar sizes and so all students judge more or less the same amount of documents. The union of all these documents conform the *pooled* document collection.

In previous years, we proceeded as follows from this point on. Although unlikely, the results provided by the pooling systems might still leave out relevant documents. To assure that all pools have some relevant material, we always included Google's top $k_G$ results for each topic, as we checked when selecting topics that some relevant web pages were included there. Also, we added $k_N$ random documents crawled from noise topics, which we created by excluding specific terms appearing in the topic set descriptions. These noise documents allowed us to check for quality in the relevance judgments, as they should all be judged not relevant for any topic. If we found students judging these noise documents as relevant, we would have an indication of possible negligence. Therefore, all pools had $k_N$ noise documents, the first $k_G$ documents retrieved by Google, and documents retrieved by the pooling systems up to a minimum of *k* documents altogether. The union of all documents in these pools conformed the *biased* document collection. This was the collection we provided students with to run and evaluate their systems.

This year we added another processing step. Evidence from EIREX 2010 and 2011 showed that documents retrieved for one topic were relevant for a different topic. Even noise topics were relevant in some cases, especially given that we tried to make topics similar to each other [Urbano et al., 2012]. For EIREX 2012 we re-indexed the pooled document collection (no noise topics yet), and formed final pools with the outputs of the pooling systems and the $k_N$ and $k_G$ documents, up to a minimum of $k^*$ documents in the pool. This was the biased collection this year, and *these* pools dictated what documents to judge. Re-indexing the collection again, we could make the final judging pools more similar to those of the students, as both their systems and the pooling systems indexed the same pooled collection. In 2010 and 2011 students indexed the biased collection, but the judging pools were formed from the complete collection.

|      |               | Topics | | Complete Collection | | | Biased Collection | | |
|------|---------------|--------|-----------|--------|------------|----------|-------|------------|--------|
| Year | Theme         | #      | Avg. words | #      | Avg. words | Size     | #     | Avg. words | Size   |
| 2010 | Computing     | 20     | 9         | 9,769  | 1,307      | 735 MB   | 1,967 | 1,319      | 161 MB |
| 2011 | Crowdsourcing | 23     | 6         | 13,245 | 1,149      | 952 MB   | 2,088 | 902        | 96 MB  |
| 2012 | Social Media  | 35     | 7         | 22,402 | 975        | 1,457 MB | 3,738 | 808        | 209 MB |

Table 2. Summary of the EIREX test collections.

Table 2 summarizes the size of the EIREX 2010, 2011 and 2012 collections. It can be seen that this year we had a much larger collection, about twice as large as previous years. In general, documents were even smaller this year, and the topics were again similar to each other. As explained in Section 3.3, this made the pooling systems retrieve documents for a topic that were not retrieved by Google for that particular topic. That is, there was a higher level of overlap between topics, and the biased collection was thus further reduced.

## 3.1. Topics

The EIREX 2012 test collection contains a total of 35 topics, all of which pertain to the *Social Media* theme we chose. All topic descriptions have a common structure (see Figure 1) with a unique id, a title and a description of what is considered to be relevant to the topic. In 2010 we kept things simple and had a generic description of relevance levels for all topics [Urbano et al., 2011b], but from 2011 we make topic-specific descriptions.

```
<topic id="2012-014">
  <title>Social media in the Arab uprisings</title>
  <relevance>
    <level value="2">The document must discuss the role of social media sites such as Facebook, Twitter or Youtube
        in the uprisings in Arab countries such as Egypt or Tunisia.</level>
    <level value="1">The document discusses the topic, but focuses on just one site or one country in
        particular.</level>
    <level value="0">The document may discuss one particular case were social media was used, but there is no global
        information.</level>
  </relevance>
</topic>
```

Figure 1. A sample EIREX 2012 topic description.

The topic titles were used as input queries to the student systems, so they can all be considered *short automatic* runs in TREC's terminology [Voorhees et al., 2005] (i.e. there is no human intervention in creating the queries from the topic descriptions). Topic titles were short this year too (see Table 2): only 7 words on average, ranging from 3 to 12. Topics 036, 037 and 038 were used as noise topics to obtain nonrelevant documents[3].

## 3.2. Documents

The *complete* document collection contains all documents returned by Google for the final set of 35 topics plus the 3 noise topics (see Table 1). A total of 22,402 web pages were crawled for all 38 topics, which account for 1,457 MB. The median size per document is 49 KB, with a mean of 67 KB. The median number of words per document is 975, with a mean of 1,765. These documents were used just as downloaded, with no postprocessing involved. However, this year we cut all documents to 256 KB to avoid very large files (in previous collections some documents were several MB long). The *biased* collection, containing only documents in the final pools (see Section 3.3), had a total of 3,738 documents, which account for 209 MB. The median size per document is 41 KB, with a mean of 57 KB. The median number of words per document is 808, with a mean of 1,361. Compared to 2010 and 2011, the collections have about twice as many documents and are about twice as big, though they actually have less textual content (see Table 2).

---

[3] Noise topics have no description in the topics file.

### 3.3. Pools

For each of the 35 topics in the collection, we ran the 11 pooling systems described in Table 3. We used various configurations of Lemur version 4.12 and Lucene version 2.9.4, which basically differed on the stemmer, the treatment of stop words, the retrieval model employed and the use of query expansion. These were basically the same configurations as in 2010 and 2011, but changing version numbers [Urbano et al., 2011b; Urbano et al., 2012].

| Id | System | Parse HTML | Stemmer | Stop words | Model | Query expansion |
|---|---|---|---|---|---|---|
| p0013 | Lemur 4.12 | Yes | Krovetz | No | Okapi | No |
| p0014 | Lemur 4.12 | Yes | Krovetz | No | Okapi | Yes |
| p0015 | Lemur 4.12 | Yes | Krovetz | Yes | Okapi | No |
| p0016 | Lemur 4.12 | Yes | Krovetz | Yes | Okapi | Yes |
| p0017 | Lemur 4.12 | Yes | No | No | Okapi | No |
| p0018 | Lemur 4.12 | Yes | No | No | Okapi | Yes |
| p0019 | Lemur 4.12 | Yes | No | Yes | Okapi | No |
| p0020 | Lemur 4.12 | Yes | No | Yes | Okapi | Yes |
| p0021 | Lucene .net 2.9.4 | Yes | No | Yes | Vectorial | Yes |
| p0022 | Lucene .net 2.9.4 | Yes | Porter | Yes | Vectorial | No |
| p0023 | Lucene .net 2.9.4 | Yes | Porter | Yes | Vectorial | Yes |

Table 3. Summary of the EIREX 2012 pooling systems.

For each topic we ran the 12 pooling systems (Google's plus the 11 in Table 3) with the complete document collection, and pooled results until at least 100 documents were joined. As Table 1 shows, 3,388 documents were put together. This was the pooled collection. The 12 systems were again run with this collection, and we joined the top $k_G$ documents retrieved by Google and $k_N$ random documents from the three noise topics. As last year, we chose $k_G=k_N=10$ documents. Then, we pooled results from the 12 pooling systems until at least 160 documents were in the pool altogether. As shown in Table 1, pool sizes ranged between 160 and 164, with an average of 161 documents. Therefore, all students judged more or less the same amount of documents. Pool depths ranged between 50 and 129, with an average 94, showing that the pooling systems tended to agree much more for some topics than for others. Note that the sum of all pool sizes is 5,623, while the biased collection contains 3,738 unique documents (66%). This indicates that several documents were retrieved for more than one topic: 1,059 were retrieved for 2 topics, 290 were retrieved for 3 topics, 66 for 4 topics, 12 for 5 topics and finally one document was retrieved for 6 topics. In 2010, the biased collection contained 97% of the total maximum [Urbano et al., 2011b], while in 2011 it contained 90% [Urbano et al., 2012]. As intended, this year we thus achieved a much higher level of overlap across topics, making it more difficult for systems to figure out which documents were crawled for which topics.

### 3.4. Relevance Judgments

We applied a cleaning process to all web pages before being displayed to the assessors, turning them into a basic black and white document to make the reading task easier. We also removed all scripts, embedded objects and HTML elements not related to page rendering. Finally, we also highlighted terms that appeared in the topic description. This preprocessing in documents to be judged seemed to improve results and helped assessors to stay focused and perform the task faster [Urbano et al., 2011c]. Assessors were able to use a basic search option, and they of course did not know whether documents were from the Google top results or noise topics. Judging took a little below 1.5 hours per assessor, so the task could be completed in one class session. Students never had access to the relevance scores, as all files were encrypted for submission back to the course instructors.

In 2010 we had two students judge every topic-document in order to measure the effect of inconsistency on relevance judgments, that is, de degree to which using one or another assessor affects the evaluation results [Urbano et al., 2011b][Voorhees, 2000]. We found that students agreed with each other to the same degree TREC assessors do, and that differences between students' judgments did not have a significant impact on the results. Therefore, for EIREX 2011 we generally had only one student judge the ~100 documents per topic [Urbano et al., 2012]. This year, we reduced the number of documents to judge from ~100 to ~90 per student, and had two students per topic. The ~160 documents per topic contained ~140 regular documents plus 10 from Google and 10 noise documents. Each student was given a random half of the ~140, plus all 10 from Google and all 10 noisy, leading to the total of ~90 per student. The 20 non-regular documents were used to check agreement and quality. Overall, we thus have much deeper judgment pools this year, from ~100 documents to ~160. As final judgments to the Google and noise documents we just randomly selected one of the judgments by the two topic assessors.

We used a 3-point relevance scale from 0 to 2: nonrelevant, somewhat relevant, and highly relevant. Documents that could not be judged due to technical problems when rendering were judged as -1 (0.8% of the times). On average, students judged 25 documents per topic as somewhat relevant and 23 as highly relevant. Of all the 700 judgments on noise documents, just one time (0.14%) was the document judged somewhat relevant and never was it judged highly relevant. Unlike last year, this year's noise topics did therefore seem to work as well as expected. Also, the average agreement between the two assessors per topic was 0.67 as measured by Cohen's kappa with equal weights, suggesting we can trust the judgments. To compare, the average agreement between assessors in EIREX 2010 was 0.417 [Urbano et al., 2011b].

## 4. Evaluation Results

All 35 topics were used to evaluate the 56 student systems (3 modules for each of the 19 student groups[4]). We used nDCG@100 (Normalized Discounted Cumulated Gain at 100 documents retrieved) as the main measure to rank systems, and AP@100 (Average Precision), P@10 (Precision) and RR (Reciprocal Rank) as secondary measures, using a 2-point relevance scale conflating the somewhat and highly relevant levels.

Table 4 reports the mean scores for each of the four measures over the 35 topics. System 01.1 obtained the best nDCG@100 score, 0.556, and therefore received the extra point this year. Unlike previous years, the top ranks were more disputed across student groups this time. As Figure 2 shows, the rankings are very similar between nDCG@100 and AP@100 (τ=0.896), but they are very different when compared to P@10 and RR. This shows that there are systems that don't return relevant documents in general, but still they return highly relevant documents towards the top of the list. This is especially true for some systems in the lower half of the ranking according to nDCG@100, which nevertheless perform quite well according to RR.

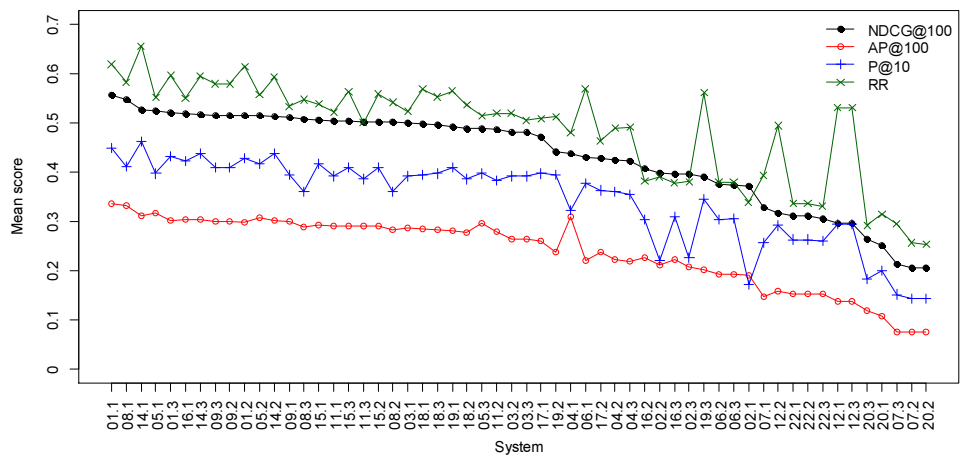

Figure 2. Mean nDCG@100, AP@100, P@10 and RR scores for the 56 student systems over the 35 topics. Systems are sorted by mean nDCG@100 score.

Once again, student systems behaved fairly well compared to usual TREC ad hoc results [Voorhees et al., 2005], probably due to the methodology followed to build the test collection (see Section 3). Documents were crawled for a prefixed set of topics, and if topics were quite different from one another (which we attempted to avoid), documents would probably be very different too. That is, it might be somewhat clear, from an algorithmic perspective, what documents pertain to what topics, although systems would still have to rank relevant documents properly. This can be observed in Figure 3. The left plot shows, for each group's best system, the average ratio of documents retrieved at different cutoffs that were crawled for each topic. Call this measure C@k (Crawl). We can see that for the best systems over 70% of the documents retrieved were actually crawled for the particular topics. The right plot displays the R@k scores (Recall), showing that the top systems retrieved over 60% of the relevant documents by the end cutoff *k*=100.

---

[4] Group 17 did not develop module 3.

| System | nDCG@100 | AP@100 | P@10 | RR |
| --- | --- | --- | --- | --- |
| 01.1 | 0.556 ± 0.176 | 0.335 ± 0.191 | 0.449 ± 0.276 | 0.619 ± 0.355 |
| 08.1 | 0.548 ± 0.169 | 0.331 ± 0.172 | 0.411 ± 0.272 | 0.583 ± 0.363 |
| 14.1 | 0.527 ± 0.164 | 0.311 ± 0.167 | 0.463 ± 0.279 | 0.656 ± 0.363 |
| 05.1 | 0.525 ± 0.184 | 0.316 ± 0.196 | 0.397 ± 0.282 | 0.552 ± 0.370 |
| 01.3 | 0.521 ± 0.183 | 0.302 ± 0.180 | 0.431 ± 0.277 | 0.596 ± 0.359 |
| 16.1 | 0.518 ± 0.171 | 0.304 ± 0.177 | 0.423 ± 0.286 | 0.551 ± 0.391 |
| 14.3 | 0.516 ± 0.173 | 0.303 ± 0.172 | 0.437 ± 0.284 | 0.595 ± 0.367 |
| 09.3 | 0.516 ± 0.180 | 0.3 ± 0.171 | 0.409 ± 0.270 | 0.579 ± 0.391 |
| 09.2 | 0.515 ± 0.180 | 0.3 ± 0.171 | 0.409 ± 0.270 | 0.579 ± 0.391 |
| 01.2 | 0.514 ± 0.189 | 0.298 ± 0.183 | 0.429 ± 0.279 | 0.614 ± 0.364 |
| 05.2 | 0.514 ± 0.185 | 0.307 ± 0.188 | 0.417 ± 0.262 | 0.557 ± 0.362 |
| 14.2 | 0.514 ± 0.176 | 0.302 ± 0.173 | 0.437 ± 0.284 | 0.593 ± 0.368 |
| 09.1 | 0.511 ± 0.180 | 0.3 ± 0.175 | 0.394 ± 0.279 | 0.533 ± 0.385 |
| 08.3 | 0.508 ± 0.179 | 0.288 ± 0.176 | 0.36 ± 0.278 | 0.547 ± 0.375 |
| 15.1 | 0.506 ± 0.169 | 0.293 ± 0.170 | 0.417 ± 0.268 | 0.538 ± 0.382 |
| 11.1 | 0.504 ± 0.177 | 0.29 ± 0.178 | 0.391 ± 0.287 | 0.522 ± 0.393 |
| 15.3 | 0.503 ± 0.171 | 0.291 ± 0.170 | 0.409 ± 0.277 | 0.564 ± 0.383 |
| 11.3 | 0.503 ± 0.179 | 0.29 ± 0.180 | 0.386 ± 0.299 | 0.5 ± 0.390 |
| 15.2 | 0.503 ± 0.174 | 0.29 ± 0.171 | 0.409 ± 0.274 | 0.558 ± 0.388 |
| 08.2 | 0.501 ± 0.181 | 0.283 ± 0.179 | 0.36 ± 0.285 | 0.541 ± 0.381 |
| 03.1 | 0.499 ± 0.168 | 0.287 ± 0.164 | 0.391 ± 0.273 | 0.523 ± 0.388 |
| 18.1 | 0.498 ± 0.168 | 0.285 ± 0.172 | 0.394 ± 0.291 | 0.568 ± 0.382 |
| 18.3 | 0.497 ± 0.177 | 0.283 ± 0.175 | 0.397 ± 0.298 | 0.553 ± 0.374 |
| 19.1 | 0.493 ± 0.153 | 0.281 ± 0.159 | 0.409 ± 0.274 | 0.565 ± 0.368 |
| 18.2 | 0.489 ± 0.174 | 0.277 ± 0.174 | 0.386 ± 0.294 | 0.536 ± 0.370 |
| 05.3 | 0.489 ± 0.221 | 0.297 ± 0.201 | 0.397 ± 0.283 | 0.514 ± 0.379 |
| 11.2 | 0.486 ± 0.191 | 0.279 ± 0.178 | 0.383 ± 0.285 | 0.519 ± 0.396 |
| 03.2 | 0.48 ± 0.161 | 0.264 ± 0.145 | 0.391 ± 0.263 | 0.519 ± 0.357 |
| 03.3 | 0.48 ± 0.161 | 0.265 ± 0.146 | 0.391 ± 0.263 | 0.505 ± 0.347 |
| 17.1 | 0.472 ± 0.165 | 0.26 ± 0.162 | 0.397 ± 0.276 | 0.509 ± 0.380 |
| 19.2 | 0.441 ± 0.177 | 0.238 ± 0.160 | 0.394 ± 0.285 | 0.512 ± 0.383 |
| 04.1 | 0.438 ± 0.174 | 0.223 ± 0.177 | 0.323 ± 0.256 | 0.48 ± 0.386 |
| 06.1 | 0.431 ± 0.168 | 0.22 ± 0.162 | 0.377 ± 0.268 | 0.569 ± 0.365 |
| 17.2 | 0.429 ± 0.207 | 0.237 ± 0.176 | 0.363 ± 0.291 | 0.463 ± 0.390 |
| 04.2 | 0.424 ± 0.179 | 0.223 ± 0.177 | 0.36 ± 0.291 | 0.49 ± 0.393 |
| 04.3 | 0.423 ± 0.174 | 0.219 ± 0.174 | 0.354 ± 0.288 | 0.491 ± 0.391 |
| 16.2 | 0.407 ± 0.225 | 0.226 ± 0.200 | 0.303 ± 0.288 | 0.381 ± 0.382 |
| 02.2 | 0.399 ± 0.237 | 0.211 ± 0.194 | 0.22 ± 0.246 | 0.39 ± 0.395 |
| 16.3 | 0.397 ± 0.226 | 0.222 ± 0.200 | 0.309 ± 0.284 | 0.378 ± 0.364 |
| 02.3 | 0.395 ± 0.240 | 0.207 ± 0.193 | 0.226 ± 0.259 | 0.381 ± 0.381 |
| 19.3 | 0.391 ± 0.168 | 0.201 ± 0.155 | 0.346 ± 0.274 | 0.561 ± 0.391 |
| 06.2 | 0.375 ± 0.201 | 0.192 ± 0.175 | 0.303 ± 0.280 | 0.38 ± 0.365 |
| 06.3 | 0.374 ± 0.200 | 0.192 ± 0.175 | 0.306 ± 0.284 | 0.38 ± 0.366 |
| 02.1 | 0.371 ± 0.234 | 0.191 ± 0.186 | 0.171 ± 0.223 | 0.339 ± 0.387 |
| 07.1 | 0.328 ± 0.184 | 0.147 ± 0.147 | 0.257 ± 0.216 | 0.393 ± 0.342 |
| 12.2 | 0.317 ± 0.218 | 0.158 ± 0.155 | 0.291 ± 0.276 | 0.494 ± 0.418 |
| 22.1 | 0.311 ± 0.205 | 0.153 ± 0.161 | 0.263 ± 0.261 | 0.336 ± 0.291 |
| 22.2 | 0.311 ± 0.205 | 0.153 ± 0.161 | 0.263 ± 0.261 | 0.336 ± 0.291 |
| 22.3 | 0.305 ± 0.213 | 0.153 ± 0.162 | 0.26 ± 0.264 | 0.33 ± 0.296 |
| 12.1 | 0.296 ± 0.188 | 0.138 ± 0.147 | 0.294 ± 0.275 | 0.531 ± 0.409 |
| 12.3 | 0.296 ± 0.188 | 0.138 ± 0.147 | 0.294 ± 0.275 | 0.531 ± 0.409 |
| 20.3 | 0.264 ± 0.199 | 0.118 ± 0.137 | 0.183 ± 0.244 | 0.292 ± 0.355 |
| 20.1 | 0.25 ± 0.186 | 0.106 ± 0.122 | 0.2 ± 0.235 | 0.314 ± 0.366 |
| 07.3 | 0.213 ± 0.154 | 0.076 ± 0.090 | 0.151 ± 0.180 | 0.295 ± 0.346 |
| 07.2 | 0.205 ± 0.164 | 0.075 ± 0.101 | 0.143 ± 0.190 | 0.256 ± 0.325 |
| 20.2 | 0.204 ± 0.151 | 0.074 ± 0.087 | 0.143 ± 0.197 | 0.253 ± 0.331 |

Table 4. Mean and standard deviation of the nDCG@100, AP@100, P@10 and RR scores for the 56 student systems over 35 topics. Systems are ordered by mean nDCG@100 score.

In EIREX 2010, most of the best systems obtained C@k scores between 75% and 100% and R@100 scores around 90%. That is, it seemed much easier to find the relevant material for each topic. In 2011, most of the best systems obtained C@k scores around 50% and R@100 scores around 60%. These differences could be attributable to the fact that both in 2011 and this year we had a larger overlap among topics (see Section 3.3), so the retrieval task is therefore harder.

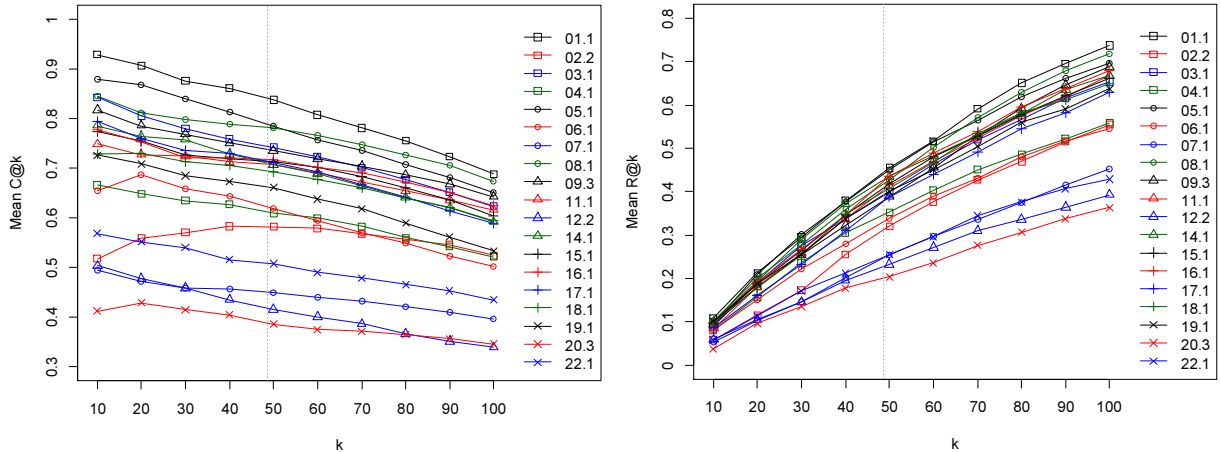

Figure 3. Mean C@k (left) and R@k (right) for the best system per student group. The grey vertical line marks the mean number of relevant documents across topics (49).

# 5. Incompleteness of Relevance Judgments

A drawback of TREC-like evaluations is that the sets of relevance judgments are incomplete, because only the documents in the pools are judged [Voorhees, 2002]. Of course, if a system does not have the opportunity to contribute much to the pool it is expected to have its effectiveness diminished, as it might have retrieved relevant material which is unknown. In the worst case, a brand new system using these collections did not contribute at all to the pools, and so its evaluation could be unreliable. This is our case, as the systems developed by the students did not contribute to the pools, only the Lemur and Lucene systems did (the pooling systems). The effect of incompleteness has been studied with TREC dada, concluding that the early ad hoc tracks were quite robust to the incompleteness problem [Zobel, 1998]. In previous years, we conducted a similar analysis with the EIREX 2010 and 2011 collections, and we also found that the small pools we used were quite reliable despite their depth differences [Urbano et al., 2011b][Urbano et al., 2012].

In this 2012 edition we further studied this issue by looking at even deeper pools. We generated pools of size 20, with the $k_N$=10 noise documents and the top $k_G$=10 retrieved by Google for each topic. Then, we added documents returned by the pooling systems to a minimum pool size of 30, 40, 50, and so on, up to the final pools of at least 160 documents. This gives us 15 different pools, each of which can be used to evaluate with the corresponding set of relevance judgments (*trel*, for topic relevance set) from Section 4. We evaluated the 56 student systems for each pool. Then, for each increment of 10 documents in the pool, we calculated the relative difference in effectiveness for each system between the two pools. The difference is measured as the percentage increased in effectiveness from the smaller to the larger pool, so that it is directly comparable with Zobel's findings (differences between 0.5% and 3.5%, with some observations of up to 19% in TREC-3). Figure 4 shows how the average difference in effectiveness diminishes as the pool size increases.

In the case of nDCG@100, pool sizes larger than 90 show subsequent increments below 1% except for one case (see Table 5). In the case of AP@100, pools with more than 90 documents correspond to relative differences between 0% and 3%, with an average of 1.5%. These results show that the pools seem again to be reliable compared to TREC's. Compared to previous years though, we see larger increments, but small nonetheless. This is especially clear in P@10, which still shows somewhat large differences with deep pools. We believe these differences in just the top 10 documents retrieved are again evidence that the task was harder this year. The increments in RR also seem to support this. Overall, it seems that going beyond size-100 does not provide enough additional reliability for the effort, though the unstable P@10 and RR clearly benefit from larger pools. In any case, the correlations between system rankings are extremely similar between pools, especially for nDCG@100 and AP@100 (see Table 5). Therefore, the relative ordering of systems, which is the main course grading criterion, is stable.

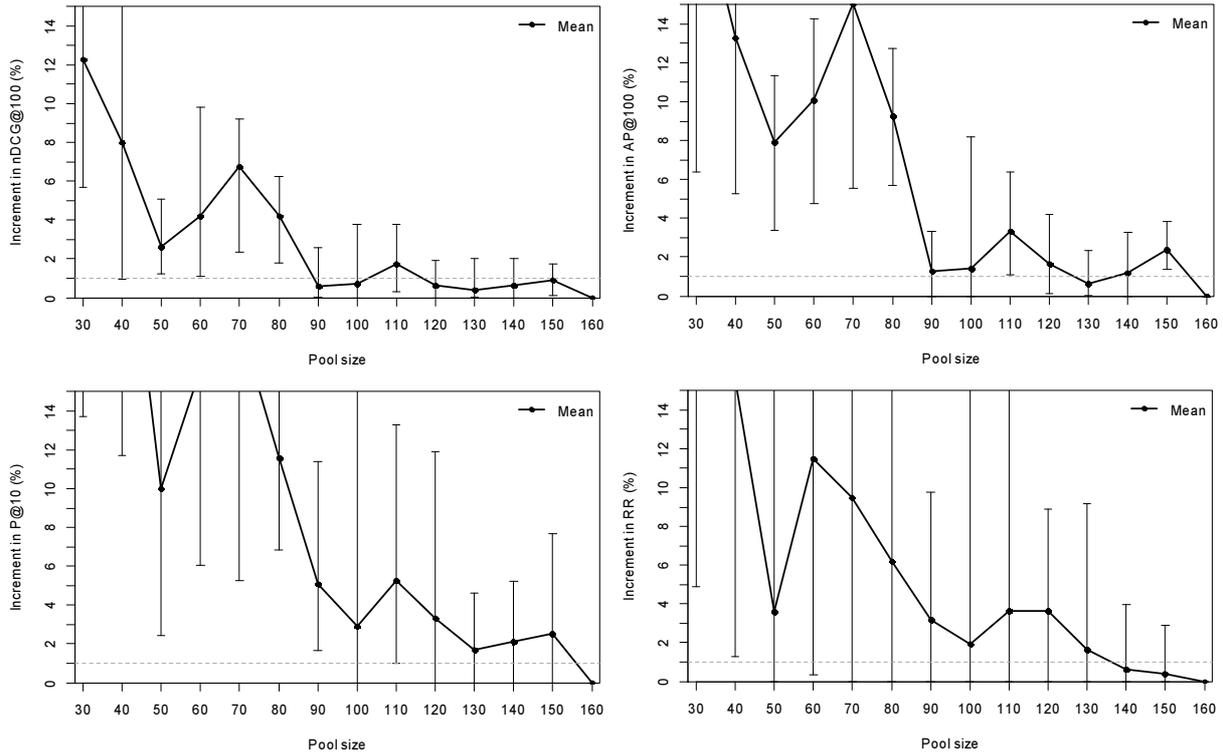

Figure 4. Mean nDCG@100 (top left), AP@100 (top right), P@10 (bottom left) and RR (bottom right) increments as a function of pool size (lower is better). Error bars show the range of increments observed.

| Pool size | nDCG@100 | | | AP@100 | | | P@10 | | | RR | | |
|---|---|---|---|---|---|---|---|---|---|---|---|---|
| | Mean | Max | τ | Mean | Max | τ | Mean | Max | τ | Mean | Max | τ |
| 20 → 30 | 12.27% | 18.74% | 0.897 | 21.49% | 37.44% | 0.878 | 31.74% | 56.24% | 0.916 | 22.25% | 52.06% | 0.719 |
| 30 → 40 | 8.01% | 16.20% | 0.911 | 13.27% | 28.83% | 0.920 | 26.55% | 48.04% | 0.912 | 15.58% | 39.32% | 0.802 |
| 40 → 50 | 2.62% | 5.08% | 0.971 | 7.90% | 11.37% | 0.949 | 9.98% | 19.37% | 0.932 | 3.58% | 15.27% | 0.885 |
| 50 → 60 | 4.22% | 9.79% | 0.942 | 10.09% | 14.25% | 0.934 | 16.11% | 26.13% | 0.922 | 11.50% | 25.70% | 0.838 |
| 60 → 70 | 6.77% | 9.21% | 0.967 | 15.06% | 19.63% | 0.954 | 18.59% | 27.18% | 0.904 | 9.51% | 26.40% | 0.841 |
| 70 → 80 | 4.22% | 6.22% | 0.952 | 9.27% | 12.73% | 0.963 | 11.57% | 29.15% | 0.936 | 6.18% | 16.21% | 0.887 |
| 80 → 90 | 0.58% | 2.58% | 0.979 | 1.27% | 3.32% | 0.969 | 5.07% | 11.40% | 0.959 | 3.20% | 9.76% | 0.935 |
| 90 → 100 | 0.72% | 3.79% | 0.987 | 1.42% | 8.17% | 0.986 | 2.93% | 15.91% | 0.948 | 1.94% | 15.82% | 0.962 |
| 100 → 110 | 1.74% | 3.80% | 0.964 | 3.31% | 6.37% | 0.976 | 5.26% | 13.29% | 0.950 | 3.66% | 29.15% | 0.928 |
| 110 → 120 | 0.64% | 1.93% | 0.973 | 1.64% | 4.20% | 0.976 | 3.30% | 11.92% | 0.955 | 3.65% | 8.91% | 0.882 |
| 120 → 130 | 0.41% | 2.01% | 0.983 | 0.63% | 2.33% | 0.980 | 1.68% | 4.64% | 0.990 | 1.67% | 9.17% | 0.945 |
| 130 → 140 | 0.62% | 2.00% | 0.979 | 1.21% | 3.27% | 0.984 | 2.13% | 5.22% | 0.989 | 0.64% | 3.98% | 0.965 |
| 140 → 150 | 0.91% | 1.73% | 0.991 | 2.39% | 3.85% | 0.984 | 2.54% | 7.70% | 0.954 | 0.39% | 2.88% | 0.992 |
| 150 → 160 | 0.00% | 0.00% | 1.000 | 0.00% | 0.00% | 1.000 | 0.00% | 0.00% | 1.000 | 0.00% | 0.00% | 1.000 |

Table 5. Mean and maximum relative increments observed in nDCG@100, AP@100, P@10 and RR, and Kendall τ correlation between the rankings, of all 56 systems, as a function of pool size.

# 6. Conclusions

In 2010 we run the first Information Retrieval Education through EXperimentation (EIREX) experiment to bring TREC-like evaluations to the IR undergraduate classroom [Urbano et al., 2011b]. After the successful experience, we repeated in 2011 with a smaller number of students, but we were able to build a larger test collection taking advantage of the reliability analysis from 2010 [Urbano et al., 2012]. In 2012 we proceeded with the third edition of the series: EIREX 2012. This year we had a much larger group of students, which allowed us to build a much larger collection and also dig deeper into the incompleteness issue. With this initiative we get students involved in the whole process of building a search engine and a test collection to evaluate it. Our goal is to introduce students in this kind of laboratory experiments in Computer Science, with a special focus on how to evaluate their systems and analyze the results.

We have described how to adapt TREC's ad-hoc methodology to build such collections for an IR course. The first main difference is that the documents in the collection are gathered after selecting the topics, and not the other way around as usual. The second main difference is related to the pools of documents to judge: the systems developed by the students cannot contribute directly to the pools to prevent cheating, and the judging effort is limited because the students cannot be asked to judge as many documents as we would want. Due to this limitation, the pools are formed differently, with the help of freely available IR tools. The question is whether such small-scale experiments are reliable or not, which is again an excellent question to investigate with the students, so they learn how to analyze them from a critical point of view to look into possible threats to validity [Voorhees, 2002][Urbano, 2011]. The main threats to validity in our case are the inconsistency and incompleteness of relevance judgments. With the EIREX 2010 experiment we found high agreement scores between students, and very high correlations between system rankings when using different sets of relevance judgments; in terms of incompleteness, we estimated that pools of size 100 and different depths were quite reliable and did not seem to affect the evaluation significantly [Urbano et al., 2011b]. In EIREX 2011 we also analyzed the effect of incompleteness, and found very similar results [Urbano et al., 2012]. This year we looked at pools of size 160, and again conclude that about 100 documents seem appropriate. We therefore conclude that the judgments made by students can be trusted, and that the pooling method proposed seems to work reasonably well for these small-scale evaluations.